\documentclass[prb]{revtex4}

\usepackage{graphicx}
\usepackage{dcolumn}
\usepackage{bm}
\begin{document}

\title{\bf Microwave induced resistance oscillations on a high-mobility 2DEG: absorption/reflection and   temperature damping experiments. }

\author{S.A. Studenikin$^1$,
M. Potemski$^{1,2}$,  A. Sachrajda$^1$, M.Hilke$^3$, L. N. Pfeiffer$^4$, K. W. West$^4$  } 
\affiliation{
$^1$Institute for Microstructural Sciences,National Research Council of Canada, Ottawa, Ontario, K1A OR6,
Canada\\
$^2$Grenoble High Magnetic Field Laboratory, MPI/FKF and CNRS, BP 166, 38042 Grenoble, Cedex 9, France \\
$^3$Department Of Physics, McGill University, Montreal, Canada H3A 2T8  \\
$^4$ Bell Laboratories, Lucent Technologies, Murray Hill, New Jersey, 07974
}
\date{ October 28, 2004 } 

\begin{abstract}

In this work we address experimentally a number of unresolved issues related to microwave induced resistance oscillations (MIRO) and the zero-resistance states observed recently on very high-mobility 2D electron gases in GaAs/AlGaAs heterostructures. 
In particular, we examine electrodynamic effects via reflection/absorption experiments and study the exact waveform of MIRO and their damping due to temperature.   
It is shown that electrodynamic effects due to metallic like reflection and  plasmons are important producing a wide cyclotron resonance line and a number of oscillations which do not coincide with the MIRO.  
To describe the MIRO waveform a simple model was employed involving radiation-induced scattering with displacement. 
A very good correlation was found between the temperature dependencies of the quantum lifetime from MIRO and the transport scattering time from the electron mobility.  
The results are compared with measurements of Shubnikov-de Haas oscillations down to 30 mK on the same sample.

\end{abstract}
\maketitle
\newpage

\section{Introduction}

Microwave-induced resistance oscillations (MIRO), originally predicted by Ryzhii et al. \cite{Ryzhii1970} and observed 30 years later on high mobility 2DEG samples by Zudov et al. \cite{Zudov}, have recently been the subject of intense study.  Interest in this phenomenon was largely stimulated by the unexpected observation of so called zero-resistance states on very high mobility samples \cite{Mani,Zudov2}. 
The experimental investigations have been accompanied by a stream of theoretical papers, e.g.  \cite{Ryzhii1,Vavilov,Durst} and references therein. To explain the MIRO, models involve microwave-induced scattering with displacement due to impurities \cite{Ryzhii1986,Durst} and phonons \cite{Ryzhii1,Ryzhii2}, a non-equilibrium oscillating distribution function \cite{Dorozhkin,Dmitriev}, or electron plasma effects \cite{Volkov,Mikhailov}. All of these effects may result in a negative local conductivity leading to an instability and the formation of current domains \cite{Vavilov,Andreev} which may manifest themselves experimentally as zero-resistance states. 
There also exist more elaborate models for MIRO which involve the formation of an energy gap at the Fermi level, analogous to superconductivity.  \cite{Mani,Fujita}
In spite of these novel theoretical ideas, there still exists no final consensus for the microscopic origin of the microwave induced oscillations and many unresolved experimental questions remain. 

In this work we examine electrodynamic aspects of MIRO \cite{Mikhailov,Kukushkin}
 by performing simultaneous absorption/reflection and {\it dc} transport experiments with microwaves (MWs) on a high mobility electron gas in a GaAs/AlGaAs heterostructure. It is confirmed that electrodynamic effects are important. The waveform obtained from reflection measurements in the MIRO regime is found to differ significantly from that acquired in {\it dc} transport. 

We also address the temperature activation of MIRO and waveform details.
A characteristic feature of MIRO is their persistence to much higher temperatures than the Shubnikov - de Haas (SdH) oscillations. \cite{Mani,Zudov2}  An activation analysis for the individual zero-state minima in Refs.~\onlinecite{Mani,Zudov2} produced surprisingly large values for the activation energies, several times larger than either the microwave photon or the cyclotron energies at the same magnetic field. 
As an alternative to examining individual minima to extract the activation energy, we analyze the exact waveform of the oscillations vs magnetic field using the theoretical model suggested by Ryzhii et al. \cite{Ryzhii1970,Ryzhii1986} and revised later by Durst et al. \cite{Durst}.  We find that in the semi-classical regime of large filling factors, this model successfully describes the waveform and temperature dependence of the MIRO amplitude.  
The shape of the microwave oscillations depends only on one fitting parameter, namely the width of the Landau levels (LLs),  therefore providing us with a unique tool to directly access the quantum scattering time. A quick estimate of the LLs width can be obtained from the exact positions of the minima/maxima of the oscillations. 
The data are compared with the Dingle damping parameter obtained from SdH measurements made at very low temperatures on the same sample.

\section{Experimental}

In this study we use a two dimensional electron gas (2DEG) formed in a triangular quantum well at the interface of a Si-modulation doped (100) GaAs/Al$_x$Ga$_{1-x}$As heterostructure with x=0.32. The undoped spacer was 800~\AA  while  the  2DEG was 1900~\AA below the surface.
For the electron transport measurements the sample was cleaved into a rectangular shape of $\sim$ 2x5 mm$^2$.  Small indium contacts were diffused at the edges by annealing in forming gas (a mixture of 10\% H$_2$ and 90\% N$_2$) at 420$^\circ$C for 10 minutes. 
 After a brief illumination with a red LED the 2D electron gas attained the concentration and mobility, $n=1.9\times10^{11}cm^{-2}$ and  $\mu \approx$ 10 $\times10^{6}cm^{2}/Vs$ respectively at 2 K.  
The microwave measurements were performed from 1.5 and 4.2 K in a He$^4$ cryostat. 
As the source of the microwaves (MW) an Anritsu, model 69377B, signal generator was used operating at frequencies up to 50 GHz with a typical output power of a few milliwatts.  The microwave radiation was delivered into the cryostat through a semi rigid, 0.085" diameter, copper-beryllium coaxial cable. 
The central conductor of the coaxial cable extended a few millimeters from the outer conductor. This served as an antenna for irradiating the sample mounted a few millimeters away.  A small cylindrical cavity made from a MW absorbing material was constructed around the sample to suppress cavity modes. Such modes can originate from  metallic parts of the cryostat.

The magnetic field normal to the sample surface was produced by a superconducting magnet. It was carefully calibrated using ESR, Hall effect probes, and the weak antilocalization effect. \cite{ourWAL}     
Longitudinal and transverse resistances of the 2DEG were measured using standard techniques with an AVS-47 Resistance Bridge.
To attain independent information about the modulation of the electron density of states and the shape of the Landau levels (LLs), low-temperature measurements of the SdH oscillations were performed on a He$_3$/He$_4$ dilution refrigerator at temperatures down to 30 mK.
During the MW absorption/reflection experiments the MW intensity was detected by a sensor based on an Allen Bradley carbon thermoresistor. The resistor was thinned down by polishing and encapsulated in a small plastic housing to reduce heat exchange with the helium bath.  A more detailed description of the geometry of the reflection/absorption experiments is provided in the following section.

\subsection{The reflection experiment}
The reflection experiments were performed simultaneously with {\it dc} transport measurements on the high-mobility sample in the regime where strong resistance oscillations occur.  These measurements were performed on a small sample ($\sim$ 2$\times$5 mm$^2$) under strong microwave excitation. 
The MW sensor was placed in close proximity to the sample (a few millimeters above it).  In this configuration the geometry of the electromagnetic field was not well enough defined to allow us to quantitatively calculate the field distribution or the reflection coefficient, but it provided us with qualitative information on the dynamic properties of the 2DEG system in the MIRO regime.

Figure 1 shows an example of the MW sensor response ($R_t$) and {\it dc} resistance ($R_{xx}$) measured simultaneously at T = 1.38 K with and without microwaves at f=49.75~GHz.   
It is evident from the figure that the waveforms seen in the reflection experiment differ from  {\it dc} transport measurements.  
  
\begin{figure}[tbp]
\includegraphics[width=68mm,clip=false]{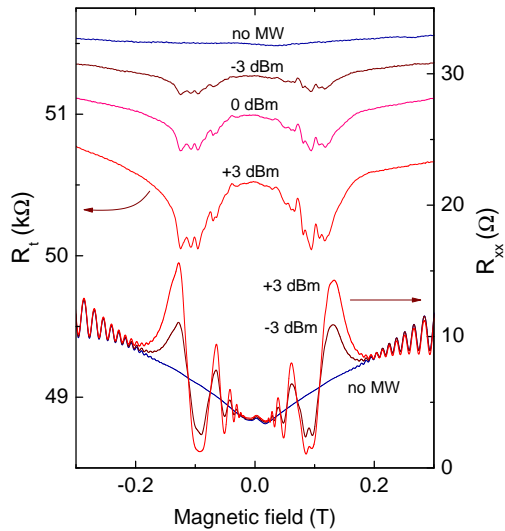}
\caption{ Experimental traces of the diagonal magnetoresistance ($R_{xx}$) and the carbon thermoresistor response ($R_t$) measured simultaneously at f=49.75 GHz at different MW powers, T=1.38 K. } 
\label{fig1}
\end{figure}

The microwave induced resistance oscillations with flat regions around $\pm$ 0.1 T corresponding to the so-called zero-resistance states (bottom traces in Fig. 1) are easy to recognize. \cite{Mani,Zudov2} These oscillations are periodic in the inverse magnetic field with a period defined by the ratio  $\omega /\omega_c$, where $\omega =2\pi~f$ is the MW circular  frequency, and $\omega _c=eB/m^*$ is the cyclotron frequency, $m^*$ is the effective electron mass.   

It is evident from Fig. 1 that the oscillation pattern in the reflection ($R_t$) measurements is very different from that observed in the transport measurements ($R_{xx}$). The position and waveform of the reflection oscillations do not depend on the MW power or temperature, but change with the MW frequency and the illumination used to vary the electron density (not shown here).  
In the reflection experiments there are two broad minima whose position roughly corresponds to the cyclotron resonance. This will be discussed in more detail below. 
The broad line is modulated with additional, faster oscillations whose positions are not matched to the  oscillations in $R_{xx}$.
There is some correlation, however, between the oscillations in $R_{xx}$ and $R_t$: both oscillate over approximately the same magnetic field range, between 0.02 and 0.2 T and cease to oscillate outside of this range.
This suggests that MIRO and the MW field oscillations are related, while revealing different patterns.
This observation may be interpreted in the following way.  The microwave radiation induces a large number of plasmon excitations leading to electron transitions between Landau levels.  However, the {\it dc} conductivity of a degenerate system in a quasi classical regime such as the one under study is not sensitive to heating, therefore, not all of the transitions manifest themselves in transport measurements, but only those that produce a large momentum transfer (i.e. a large displacement) due to disorder leading to MIRO.  

It is important to note that oscillations of the MW sensor $R_t$ occur continuously throughout the whole MIRO regime without any noticeable change in the flat areas, i.e. the zero-resistance states.  If the zero-resistance states were due to a superconductivity effect \cite{Mani,Fujita} or any similar phase-transition phenomenon, one would expect to observe a dramatic change in the reflectivity. This is not observed in Fig.~1.

One proposed explanation for the zero-resistance state involves the formation of current domains resulting from a negative local conductivity. \cite{Vavilov,Andreev}
While domain formation maybe an important element in the explanation of the apparent zero-resistance state, it is clear that the domains do not reveal themselves in the reflection experiment because, as noted above, $R_t$ oscillates throughout the whole MIRO regime and not only in regions of the zero-resistance states where domains are expected to form.  Further experimental studies, e.g. to observe domain formation in the MIRO regime,
would be beneficial. 
It is known that magnetoplasmon effects may play an important role in MW experiments on finite size samples.\cite{Mikhailov,Klitzing1993,Pan2003,Kukushkin2003} 
It is likely, that the MW field oscillations detected by $R_t$ in Fig.~1 are caused by bulk and edge plasmons due to the finite size of the sample.  Contacts along the edge of the sample may also aid the formation of plasmon modes.  A detail analysis of the oscillation pattern in the reflection data is outside of the scope of this paper. 

A characteristic MIRO property is the slow decay of the amplitude of consecutive harmonics.  In other words, the amplitude of the higher MIRO harmonics is comparable to the first harmonic peak {\it j}=1.
One may ask, therefore, whether MIRO involve resonant absorption at the cyclotron resonance harmonics?  
We investigate this question experimentally in the following section.

\subsection{The absorption experiment}
In this section we describe absorption measurements where we search for evidence of absorption at harmonics of the CR,  $\omega = j\omega_c$  with $j\geq 2$.  
Our goal was to perform a quantitative measurement of the CR line shape. Care was taken, therefore, to carefully characterise the experimental geometry.
A larger sample was used to exclude the magneto-plasmon shift of the cyclotron resonance due to a finite size sample \cite{Klitzing1993,Pan2003,Kukushkin2003}.  This experiment was not, therefore performed on the small sample described in the previous section.  A simple repositioning of the MW detector ($R_t$) by  placing it behind the sample did not qualitatively change the picture shown in Fig. 1, but did change the specific oscillation pattern.

For the absorption experiment we used a sample from a second wafer with $\sim$9$\times$9 mm$^2$ dimensions.  Although this sample came from the same source (Bell Labs) with similar nominal growth parameters it had a lower electron mobility.
The electron concentration was 1.8x10$^{11}$ cm$^{-2}$, and the mobility was 1.4$\times$10$^6$ cm$^2$/Vs and did not change after illumination.  

In order to eliminate possible edge effects a MW absorbing mask with a $\sim$3 mm diameter hole was placed just behind the sample in front of the MW sensor. Thus only radiation passing through the central part of the sample was detected.
Finally to make modelling possible we were careful to achieve a transverse electromagnetic field geometry. For this purpose the antenna was placed approximately 30 mm away from the sample.  Measurements were performed using small MW powers.   
For this geometry, we can apply classical Maxwell's electrodynamics to estimate the absorption by the 2DEG. The $ac$ conductivity of 2DEG is given by the following Drude equation for circularly left/right polarized radiation {\it \^E} = $E_0  \exp(\pm i\omega t)$:
\begin{equation}
\sigma ^{\pm}(\omega,B)=\frac{e n \mu}{1-i(\omega \pm \omega_c)\tau}
\label{eq1}
\end{equation}
where {\it e} is the electronic charge, $\tau$ is the transport relaxation time, $\mu=e\tau/m^*$ is the electron mobility, {\it n} is the 2DEG density ,  $\omega$ and $\omega_c$ are the MW and cyclotron frequencies respectively, $i=\sqrt{-1}$ is the imaginary unit.
\begin{figure}[tbp]
\includegraphics[width=68mm,clip=false]{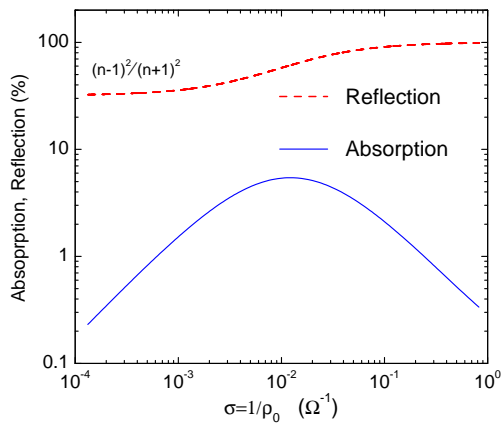}
\caption{Calculated values of absorption and reflection in per $\sigma =1/\rho_0$ at $w=w_c$. The reflection was calculated from $R=1-T-A$, where {\it T} is transmission. 
 } 
\label{fig2}
\end{figure}
We consider the 2DEG as an infinitely narrow conducting layer at the boundary between the sample and air.
Linearly polarized radiation {\it \^E} = $\frac{1}{2}E_0\bigl( \exp(-i\omega t) + \exp(i\omega t)\bigr)$, propagates along the magnetic field direction perpendicular to the {\it x-y} 2DEG plane.
 The absorbed power is given by 
$\Delta P_{\text{a}}$=$<$Re({\it \^E$^{*}$}$\sigma(\omega)${\it \^E)}$>$. 
It should be noted that the electric field {\it \^E} does not remain constant as a function of magnetic field during the sweep due to the strong reflection by the 2DEG.
As a result, the observed CR line is much broader than that given by eq.~1.
The following expression can be easily derived for the ratio of the absorbed to incident power  $\Delta P_{\text{a}}/P_i$  for linearly polarized radiation:  
\begin{equation}
	\frac{\Delta P_{\text{a}}}{P_i} =\sum_{\pm} \frac{Re(\bar \sigma^{\pm})}{(1+\kappa+Re(\bar \sigma^{\pm}))^2+(Im(\bar \sigma^{\pm}))^2}
	\label{eq2}
\end{equation}
%
where  $\kappa$ is the refractive index (3.6 for GaAs in our calculations),  
$\bar \sigma^{\pm} = \sigma^{\pm} Z_0$ is the normalized sheet conductivity of the 2DEG, and  $Z_0=\sqrt{\frac{\mu_0}{\epsilon_0}}$=377 $\Omega$ is the impedance of free space.
Higher order corrections due to interference effects were not taken into account.  

The absorption results as a function of sheet conductivity, described by eq.2, are plotted in Fig.~2. It is evident that even within this classical approach electrodynamic effects are important.  Only a very small amount of the MW power is absorbed by the high mobility 2DEG due to a large impedance mismatch between the sample conductivity and the vacuum. A very small part of the electromagnetic field penetrates into the sample. 
It is seen from the figure that for high conductivity samples ($\sigma_0 \geq 4/Z_0$) the absorption decreases with increasing conductivity.  For our experiments the sample conductivity was 0.3 and 0.04 $\Omega$ and the maximum absorption at the resonance was 0.8 \% and 3.8 \% respectively.  

For low conductivity samples one would expect two sharp CR peaks, for linearly polarized radiation, at $\omega_c = \pm \omega$, plotted in Fig.~3 by a thin solid line (5). 
The full width at the half maximum (FWHM) is inversely proportional to the electron mobility, FWHM=2/$\mu$, giving 2 and 14 mT for the samples described above. 
However, in the experiment on a high-mobility sample in Fig.~3 the CR line was much broader. The experimental FWHM of the CR absorption line in Fig.~3 is 70 mT that is 5 times broader than predicted by eq.~(1). For higher mobility samples the difference would be even larger. 

The broad maximum in Fig. 3 at B=0 is related to a parasitic B-dependent sensitivity of the MW detector and can be disregarded.  A small detector related magnetoresistance was noticeable in the small signal regime in the absorption experiments (n.b. the temperature bar on Fig. 3) but was not important in the MIRO regime at large MW powers and lower temperatures.  The small asymmetry of the experimental curves in Fig. 3 is due to a small temperature drift caused by microwaves.

\begin{figure}[tbp]
\includegraphics[width=68mm,clip=false]{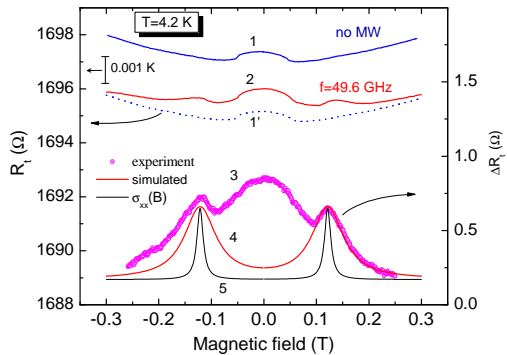}
\caption{Absorption by the high mobility 2DEG vs magnetic field at {\it f}=49.7 GHz, and T=4.2 K. \\
Lines 1 and 2 are magnetoresistance of the MW detector with and without the microwaves applied; 
curve 3 is the relative change of the MW sensor response due to the transmitted MW radiation proportional to the absorption obtained by subtracting the background dependence (curve 1');  curve 4  is the theoretical dependence from eq.~(2) in arbitrary units ;  curve 5  is the real part of the dynamic conductivity $\sigma_{xx}(\omega)$ from eq.~(1) expected for a low conductivity sample.
 } 
\label{fig3}
\end{figure}

The solid line 4 in Fig.3 is the theoretical curve based on eq.(2) using the transport parameters for this sample. 
It is clear that the simulation reproduces  the experimentally observed width and shape of the CR peak quite well.  The calculated shape of the CR (curve 4) is very close to being Lorentzian.

There is one further conclusion that may be drawn from the absorption experiment with  respect to the question of whether there is a resonant absorption at the CR harmonics.
From Fig.~1 we note that consecutive MIRO peaks in $R_{xx}$ have comparable amplitudes.   
If there existed similar absorption at each of the CR harmonics,  it would resul in sharp peaks in the absorption experiment.  Broadening would be less important for the harmonics because of the smaller partial {\it ac} conductivity, $\sigma_{j\omega}$, and these peaks would be easily detected in the absorption experiment.  

From this experiment we conclude that, at least for the lower mobility sample, there was no resonant absorption at the harmonics of the CR. Therefore, it is likely that non-resonant absorption is responsible for the MIRO. \cite{Ryzhii1986,Durst}  In this model  the impurity/phonon assisted, non-resonant absorption occurs continuously throughout the B-sweep. The MIRO occur due to the combined effect of the microwaves and the stationary magnetic and electric fields forcing electrons to predominantly scatter along or against the $dc$ bias.  
It should be noted, that there is another theoretical approach \cite{Dorozhkin,Dmitriev} in which the MIRO are explained by the MW induced change of the distribution function.  However, to a first approximation these two approaches produce very similar final equations for the conductivity \cite{Dmitriev} and at the moment it is difficult experimentally to distinguish between the two theories.

\subsection{Modulation of the electron density of states}

\begin{figure}[tbp]
\includegraphics[width=68mm,clip=false]{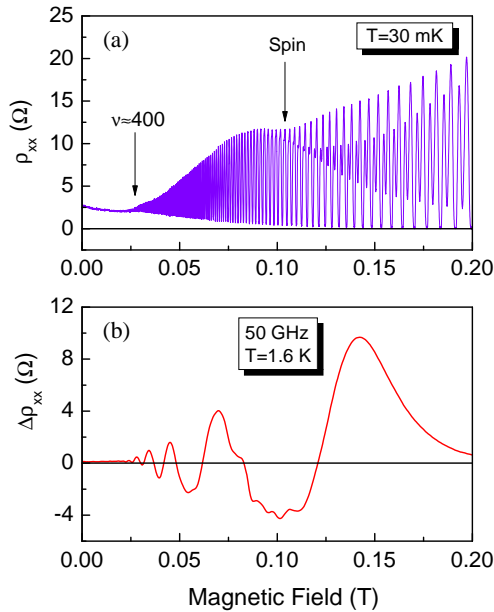}
\caption{ (a) Shubnikov -de Haas oscillations at 30 mK on a high mobility GaAs/AlGaAs sample;  \\
(b)  microwave-induced oscillations in $\Delta \rho_{xx}$ on the same sample at {\it f}=50 GHz and T=1.6 K.  
 } 
\label{fig4}

\end{figure}
\begin{figure}[tbp]
\includegraphics[width=68mm,clip=false]{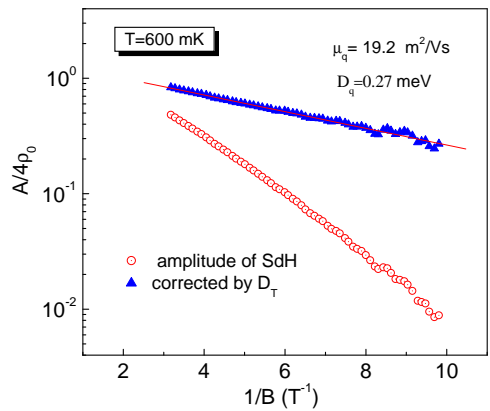}
\caption{Normalized amplitude of the SdH oscillations (open dots), and the amplitude corrected by the temperature damping factor  $D_{T} = X_T /\sinh(X_T)$ in eq. 3 (solid triangles) for T=600 mK.}  
\label{fig5}
\end{figure}

An interesting feature of MIRO is their persistence to relatively high temperatures at which SdH oscillations are totally damped.    
A number of theoretical models explicitly assume a strong modulation of the electron density of states (DOS) \cite{Durst,Ryzhii1}, other semiclassical models e.g. \cite{Vavilov} assume that the electrons traverse many orbits before scattering.   
To obtain experimental information on the strength of the DOS modulation and Landau level widths in the MIRO regime we performed magneto-transport measurements on the same sample at very low temperatures in a He$^3$/He$^4$ dilution refrigerator. 
  
Figure 4 (a) shows a trace of the SdH oscillations at 30 mK on the sample used for the reflection experiments in Fig.~1.  Figure 4 (b) shows the microwave-induced oscillations in the longitudinal differential magnetoresistance, $\Delta\rho_{xx}=\rho_{xx}^{MW}-\rho_{xx}^0$,  at 1.6~K on the same sample after a similar illumination.  
It should be mentioned that this sample also revealed microwave-induced oscillations in the transverse (Hall) component of the resistivity $\Delta\rho_{xy}$ similar to those reported in Refs. \onlinecite{ourMW} and  \onlinecite{ManiRxy}. Oscillations in $\Delta\rho_{xy}$  may serve as a useful feature to test different theoretical models of MIRO. \cite{Durst,Dmitriev,RyzhiiRxy}  

It is evident from Fig. 4 that at low temperatures the SdH oscillations persist to very small magnetic fields (filling factors up to $\nu\sim$400), and that both SdH  and microwave-induced oscillations commence at the same magnetic field around 0.025 mT. Although in the current experiment we deal with the quasi-classical situation of large filling factors, it can be qualitatively concluded  that a strong DOS modulation is an important factor to observe MIRO.  On the other hand, the two oscillations have quite distinct temperature dependences. This is the next issue we address.

For the analysis of the SdH oscillations we use the following well known expression for the oscillatory component where only the first harmonic of the Fourier expansion remains:\cite{Lifshitz,Isihara,Coleridge}:
\begin{equation}
\Delta \rho_{xx} = 4 \rho_0 D_{T}(X_T) \exp(-D_q/\hbar \omega_c)\cos (\frac{2\pi \epsilon}{\hbar \omega_c}-\pi), 
\label{eq.3}
\end{equation}
where $\rho_0$ is the zero field resistivity, $k_B$ is the Boltzman constant, $D_{SdH}=\pi \hbar/\tau_q$ is the Dingle damping parameter,  $\tau_q$ is the quantum lifetime, 
$D_{T} = X_T /\sinh(X_T)$ is the thermal damping factor with $X_T = 2\pi^{2} k_BT / \hbar\omega_{c}$.
The above expression assumes that Landau levels  have a Lorentzian shape $D_0(E)=\frac{1/\pi\Gamma}{1+(E/\Gamma)^2}$  with a full width at a half maximum FWHM=$2\Gamma$, where $\Gamma = \hbar/2\tau_q$.  
In the low-temperature experiment shown in Fig.~4 we found that the amplitude of the SdH oscillations {\it vs} magnetic filed is well described by eq.~3.
Temperature dependences of the amplitude at different fields on the illuminated sample were also fitted well  by eq.~(3)  using the correct value of the effective mass $m^*/m_0$=0.067.  These facts are strongly indicative that the LLs possesed a Lorentzian shape.  We stress this is true for large filling factors on an illuminated sample. The situation may well be different in other samples.

It is interesting to note, that the argument in eq.~3 in the thermal damping factor  is $2\pi^2$ times larger than $k_BT/\hbar \omega_c$ which makes the SdH oscillations decay very fast with temperature.   
This factor of $2\pi^2$ arises from the temperature broadening of the Fermi function and the Lorentzian shape of the LLs.

Figure 5 shows an example of the Dingle plot of the normalized amplitude of the SdH oscillations (open circles) $vs$ 1/B at T=600 mK.   Full triangles present the amplitude corrected by the thermal factor $D_{T}$  in eq.~(3).   
It is seen that the corrected amplitude $A/4\rho_0 / D_T$ is a straight line in a semi-logarithmic plot which intersects the ordinate axis close to 1.0 as expected from eq.(3), which may  not always be the case. \cite{Coleridge2} 
This is another indication that LLs possessed a Lorentzian shape.  
From the plot in Fig.~5 we deduced a quantum mobility $\mu_q=e\tau_q/m^*$=19.2 $m^2/Vs$.  Equivalently, it can be expressed in terms of the width of individual LLs $\Gamma_{SdH}$=0.043 meV, or as a Dingle damping factor in energy units $D_{SdH}=\pi \hbar/\tau_q=2\pi \Gamma_{SdH}$=0.27 meV.  We will need these numbers for comparison with similar results from MIRO.

Finally we note that our numerical simulations reveal that the amplitude of SdH oscillations $vs$ the magnetic field depend on the LL shape.  For example, if the LLs have a Gaussian shape, the Dingle plot would have a $1/B^2$ decay rather than the linear behavior observed in Fig.~5.  \cite{Coleridge2}
For large filling factors (up to 400 in Fig. 4) it is reasonable to assume that the quantum lifetime does not depend on the magnetic field. In this situation, it is reasonable to use the experimental agreement with eq.~3 to determine whether LLs have a Lorentzian shape or not. We are not aware of an explicit theoretical confirmation for this statement and it is not a trivial task to experimentally extract the exact shape of the LLs.

\subsection{The exact waveform, the phase and the temperature damping of MIRO}
As mentioned above, SdH oscillations vanish fairly rapidly with increasing temperature because of the large     $2\pi^2$  coefficient  in the thermal damping factor in eq.(3).  In contrast to SdH oscillations the MIRO persist up to much higher temperatures.  Surprisingly, the activation energies extracted from an individual MIRO minima were found to be in the 10 to 20 K range \cite{Mani,Zudov2} several times larger than either the MW photon or the cyclotron energies.  
In an alternative approach suggested in \cite{StudenikinIEEE} MIRO were fitted by a damped sinus function and reasonable values for the damping activation energy were found $D_M \sim k_BT$, which did not exceed the
relevant parameters of the experiment – i.e. the thermal, microwave, or cyclotron energies. 
The sinus wave description worked well for higher harmonics but was not successful in describing the shape and position of the first peaks.  In addition, the damping parameter $D_M$ obtained in this way varied with temperature about twice as rapidly as the electron mobility. This question needs to be resolved.

In this paper we achieve precise  description of the MIRO waveform over the whole field range.   
To describe the waveform we employed the radiation-induced scattering model. \cite{Ryzhii1986, Durst}
In this model the photon-assisted scattering rate is proportional to the product of the initial and final densities of states:
$\jmath_\pm \propto \nu (\varepsilon) \nu (\varepsilon + \hbar \omega \mp eE_{dc} \Delta x)$, where $\varepsilon$ is the  electron energy, $E_{dc}$ is the  electric field produced by $dc$ bias, $\Delta x$ is the cyclotron orbit displacement due to scattering.  It is assumed that the scattering process does not change the electron energy. The photocurrent is proportional to the difference between the scattering rates in the two directions along and against the applied bias:
$j_{ph} \propto \mathcal{M} e \Delta x \nu (\varepsilon)\partial_{\varepsilon}\nu (\varepsilon + \hbar \omega) E_{dc}$,  where  $\mathcal{M}$ is the matrix element of the transition. In general, $\mathcal{M}$ may be a function of many variables, i.e. electron and microwave energies, displacement, and the magnetic field.
It should be noted that only the fundamental cyclotron harmonic is allowed $\omega  =\omega_c$ in the dipole approximation.  However, disorder violates this strong selection rule and transitions between all LLs become allowed.~\cite{Vavilov, Durst} 
Plasmon effects, discussed above, may facilitate this violation further.
In our simulations we assume that $\mathcal{M}$=\texttt{const} meaning that the photo-assisted scattering has the same probability for all harmonics. 
In this model the MW induced photocurrent $j_{ph}$ is proportional to the bias voltage. It therefore reveals itself as a change in resistance in the Hall-bar geometry or a change in conductance in the Corbino-disc geometry. \cite{Yang-Corbino}

Based on the above assumptions and taking into account finite temperature effects through the Fermi distribution function
$n_F(\varepsilon)=1/\lbrack 1+\exp (\frac{\varepsilon - \varepsilon_F}{k_B T})\rbrack  $ 
we arrive at the following equation for the MIRO: 

\begin{equation}
\Delta \rho_{xx}(B) = A \int d\varepsilon \lbrack n_F(\varepsilon)-n_F(\varepsilon + \hbar \omega)\rbrack \nu (\varepsilon)\partial_{\varepsilon}\nu (\varepsilon + \hbar \omega)  
\label{eq.4}
\end{equation}
where {\it A} is a scaling coefficient which is field independent.   
The dependence on magnetic field comes through the density of states 
$\nu (\varepsilon)=\sum\frac{eB}{\pi \hbar} \frac{1}{\pi\Gamma  \lbrace 1+\lbrack\varepsilon -(i+1/2)\hbar \omega_c\rbrack^2/\Gamma^2\rbrace}$
where the sum is taken  over  the Landau levels index {\it i} from zero to infinity, $\Gamma$ is the LLs width.

According to eq. (4) the MIRO shape depends on a single fitting parameter $\Gamma$. This allows us to directly access the LLs broadening which is itself  inversely proportional to the quantum lifetime.

\begin{figure}[tbp]
\includegraphics[width=68mm,clip=false]{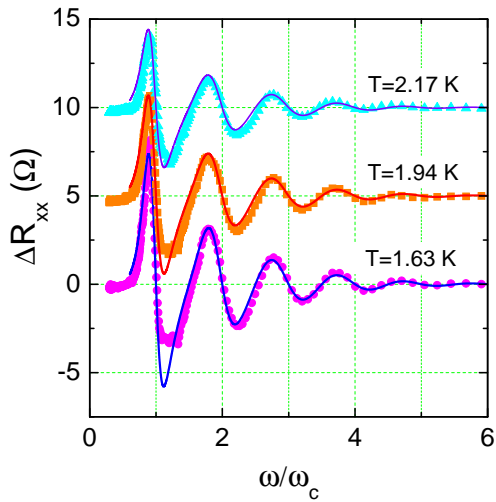}
\caption{ Microwave-induced oscillations in resistance at different temperatures.  Solid lines are the best fits with Eq. 4.}  
\label{fig6}
\end{figure}
Figure 6 shows an example of the experimental data for three different temperatures plotted as a function of the normalized inverse magnetic field $\omega / \omega _c$.  The solid lines are the best fits using Eq.(4) with the following parameters $\Gamma$=21.8, 21.2, and 20.3 $\mu eV$ for temperatures 2.17, 1.94, and 1.63~K, respectively.  
It is evident from the figure that Eq.~(4) fits the experimental waveforms well over the whole magnetic field range including the region near the first CR harmonic where there exists a large deviation from a sinusoidal function. \cite{StudenikinIEEE} We were able to fit equally well data from the much higher mobility sample ($\mu=25\times 10^6 cm^2/Vs$) available in literature. \cite{Zudov3} 

Let us discuss briefly the coefficient $A$ in eq.~4. 
When fitting data in Fig.~6 we used a constant factor $A$ for every individual curve $vs$ magnetic field, but we needed to use different values of $A$ for each temperature.  If $\nu$ in eq.~4 is calculated in units $10^{11}$~ cm$^{-2}$ and $\epsilon$ in $meV$, we obtain $A$= 11.0, 13.4, and 16.0 for temperatures 2.17, 1.94, and 1.63~K, respectively, which follows a $1/T^2$ function within an experimental uncertainty.   
Qualitatively, the temperature dependence of $A$ could be interpreted as due to the change  of the background scattering with temperature. As was discussed above there are many electron transitions induced by the microwaves.   The conductivity of a degenerate electron gas at zero or small magnetic fields is not sensitive to temperature, therefore, only  specific transitions involving large displacements (momentum transfers) are important for MIRO.  In very high mobility samples a small number of such transitions can  lead to a large effect as compared to the dark resistance, $\rho_{xx}(0)$. 
However,  number of background scattering events increases with increasing temperature which reduces probability of scattering events with large displacement. This leads to reducing MIRO amplitude controlled by coefficient $A$ in eq.~(4).

Let us turn now to discussion of the MIRO phase related to exact minima/maxima positions.\cite{Mani,Zudov2,Zudov3,Mani3}
According to theoretical predictions, e.g. see ref.\onlinecite{Ryzhii1986,Durst}, there should be no resistivity change at the exact points of the CR and its harmonics $\omega/\omega_c=j$ with {\it j} being a positive integer. To a first approximation, valid for higher harmonics, the oscillations are described by a sinus function $\Delta \rho_{xx}\propto -\sin (2\pi \omega/\omega_c)$ \cite{Durst} which gives minima/maxima positions at $\omega/\omega_c=j\pm 1/4$, respectively. (To be more precise, to a first approximation the MIRO are described by an exponentially damped sinusoid. \cite{StudenikinIEEE})
However, in experiments on very high mobility samples~\cite{Zudov3,Mani3}
 there exists a deviation from the $\pm 1/4$ phase which is evident in Fig.~6. 

Within the  model the phase shift is caused by a strong modulation of the density of states (see Fig.~4) which also leads to a strong deviation of the waveform from a simple sinusoidal function.   This distortion of the waveform appears as a phase shift of the minima/maxima from  $\pm 1/4$.  
Zudov \cite{Zudov3} obtained an analytical expression for the minima/maxima position using the simplified assumption \cite{Durst} that MIRO are proportional to the derivative of the density of states $\Delta \rho_{xx} \propto \frac{\partial \nu (\epsilon)}{\partial \epsilon}\vert_{\epsilon=\hbar \omega_c}$ taken at  $\epsilon = \hbar \omega$. 

We  used eq.~(4) to calculate the maxima/minima positions. To our surprise, the numerical simulations based on eq.~(4) show that the minima/maxima position, within an uncertainty of 1\%, obeyed exactly the same equation (eq. (8) in ref.\onlinecite{Zudov3}), but with the argument multiplied by a factor of exactly 2.  For the sake of convenience we rewrite this equation in our notation:
%
\begin{equation}
\Phi_{max/min}=\mp \frac{1}{2\pi}\arccos (\psi),\label{eq.5}
\end{equation}
where $\psi =1/2-y+\sqrt{y^2-y+9/4}$, and $y=\cosh^2(\frac{2\pi \Gamma}{\hbar \omega_c})$
We do not know if it is an exact solution of the problem based on eq.~(4), but it is a good approximation and may be used to analyze the maxima/minima positions  and to extract the LLs broadening without the need to exactly fit the waveform.
\begin{figure}[tbp]
\includegraphics[width=75mm,clip=false]{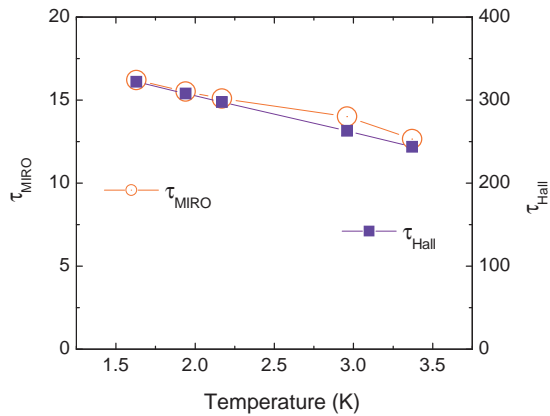}
\caption{ Transport relaxation time, $\tau_{Hall}= \mu_{Hall} m^*/e$ (right scale) determined from the Hall effect measurements, and the quantum relaxation time $\tau_{MIRO}=\hbar/2\Gamma_{MIRO}$ (left scale) determined from the MIRO.
}  
\label{fig7}
\end{figure}

Let us now discuss the temperature damping of MIRO.  
First of all, our numerical simulations reveal that the MIRO amplitude from eq.~(4) was not sensitive to the temperature broadening of the Fermi function in agreement with Durst et al. \cite{Durst}.  
Naively, this is a counterintuitive result since for a two-level system one would expect that the amplitude would be proportional to the difference between emitted and absorbed MW photons, therefore it would decay exponentially with the temperature as $\Delta \sigma \propto \lbrack 1-\exp (\hbar \omega /k_BT)\rbrack$ due to the population equilibration at higher temperatures between the two levels. \cite{Ryzhii3}
In a system with an infinite number of levels the saturation can never be reached and the total difference between up and down transitions remains constant $vs$ temperature leading to the slow, non-exponential damping of  MIRO $vs$ temperature. In this case all the temperature damping of MIRO merely comes from the temperature dependence of the quantum relaxation time.\cite{Durst}  

In order to examine this issue quantitatively we plot two scattering times in Fig.~7, the transport relaxation time, $\tau_{Hall}= \mu_{Hall} m^*/e$   determined from the Hall effect measurements, and  $\tau_{MIRO}=\hbar/2\Gamma_{MIRO}$ determined from MIRO.  It is evident  from Fig.~7  that there is a very good correlation in the relative dependences of $\tau_{Hall}$ and $\tau_{MIRO}$. The absolute values of the two relaxation times differ by a factor of 20 which is consistent with the ratio between transport and quantum relaxation times in high mobility GaAs/AlGaAs samples.\cite{Coleridge2}  This observation supports our earlier assumption in eq.~(4) that MIRO depend on the quantum relaxation time. 

Since MIRO provide us with the quantum relaxation time, let us compare this value with that obtained from the SdH oscillations shown in Fig.~4.
\begin{figure}[tbp]
\includegraphics[width=68mm,clip=false]{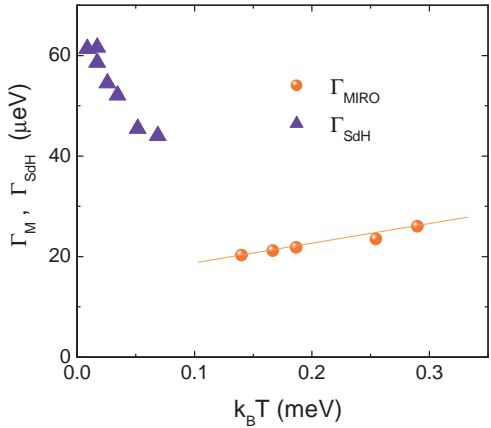}
\caption{ Width of the Landau levels extracted from MIRO ($\Gamma_{MIRO}$) and from low-temperature SdH effect measurments ($\Gamma_{SdH}$). 
}  
\label{fig8}
\end{figure}
In Fig. 8 the solid triangles represent the LLs width obtained from an analysis of the low-temperature SdH measurements using eq.~(3) and $\Gamma_{SdH} =\hbar /2\tau_q$.  The open circles represent the LLs width from MIRO.
Because of the high mobility, the SdH effect could be employed to determine $\Gamma_{SdH}$ only over a limited temperature range up to 600 mK.  At higher temperatures the thermal factor dominated the dependence in eq.~3.
Note that the quantum damping parameter, $D_{SdH}=\pi \hbar/\tau_q$ is frequently replaced by an equivalent Dingle temperature parameter $T_q$ with $D_{SdH}=2\pi^2k_BT_q$.

To our surprise we found that $\Gamma_{SdH}$ in Fig.~8 was not a constant value $vs$ temperature. The transport mobility did not change over the temperature range between 30 and 600 mK, therefore one would expect that $\tau_q$ and $\Gamma_{SdH}$  would also remain constant. 
While $\Gamma_{SdH}$ varied somewhat with the temperature, it was on average still several times larger than the value of $\Gamma_{MIRO}$ obtained from MIRO by extrapolating the dependence in Fig.~8 to zero temperature.
The observed effects of the temperature dependence of $\Gamma_{SdH}$ and the difference between $\Gamma_{SdH}$ and $\Gamma_{MIRO}$ can be qualitatively explained by the different microscopic origins of these effects.  
MIRO are a quasi classical phenomenon requiring only a modulation of the density of states and do not depend on the electron concentration or the distribution function.  In contrast to MIRO, the SdH oscillations are essentially a quantum effect depending on the LLs width, the Fermi energy, and the distribution function.  Therefore, even a very small fluctuations of the electron density over the sample surface (nb. there are very large filling factor in Fig.~4) may result in an additional damping mechanism for the SdH effect effectively increasing $\Gamma_{SdH}$ compared to $\Gamma_{MIRO}$ .

\section{conclusion}
We have experimentally studied microwave induced resistance oscillations on a high mobility GaAs/AlGaAs heterostructure under  quasi-classical conditions of large filling factors.  We found that a simple model by Ryzhii et al. and Durst et al. worked well to accurately describe the waveform and temperature damping of the MIRO. 
The temperature dependence of MIRO originates from the scattering mechanisms that results in a much slower function as compared to the exponential thermal factor in SdH effect.  The MIRO provide us with a unique tool to directly access the quantum relaxation time.  
The deviation of the minima/maxima position from $\pm$1/4 values is merely due to the large modulation of the density of states in high mobility samples in agreement with the low-temperature SdH effect measured on the same sample. 
The reflection and absorption experiments confirm the importance of  electrodynamic effects. A broad cyclotron line was observed in the absorption experiment in agreement with a classical theoretical description. 
In the reflection experiments in MIRO regime a large number of oscillations was observed due to plasmon effects as may be expected in a finite size sample.  Plasmon effects did not reveal them self in $dc$ transport measurements on very high mobility samples. 

\subsection{Acknowledgements}
We thank D.G. Austing  for his interest in this work, P. Zawadzki, M. Pioro-Ladriere, and M. Ciorga for the help in the SdH experiment.  
S.A.S and A.S. acknowledge support of The Canadian Institute for Advanced Research (CIAR).
This work was also supported by the NRC-Helmholtz joint research program.

\end{document}